# COSMOLOGICAL INFLATION

# AND

# THE NATURE OF TIME

D.S. SALOPEK

*Department of Physics*
*University of Alberta*
*Edmonton, Canada T6G 2J1*

**ABSTRACT**

Recent advances in observational cosmology are changing the way we view the nature of time. In general relativity, the freedom in choosing a time hypersurface has hampered the implementation of the theory. Fortunately, Hamilton-Jacobi theory enables one to describe all time hypersurfaces on an equal footing. Using an expansion in powers of the spatial curvature, one may solve for the wavefunctional in a semiclassical approximation. In this way, one may readily compare predictions of various inflation models with observations of microwave background anisotropies and galaxy clustering.

# 1  Introduction

In the standard cosmological formulation of Einstein's theory of gravity, one ordinarily solves the field equations for the 4-metric $g_{\mu\nu}$. In a more elegant although equivalent approach, it has been recommended [1]-[6] that one solve the Hamilton-Jacobi (HJ) equation for general relativity which governs the evolution of the generating functional $\mathcal{S}$. By adopting this method, many advantages have be gained:

**(1) Avoiding Gauge Problems.** Neither the lapse $N$ nor the shift $N_i$ appear in the HJ equation. The generating functional depends only on the 3-metric $\gamma_{ij}$ (as well any matter fields that may be present). Hence, the structure of the HJ equation is conceptually simpler than that of the Einstein field equations. For instance, one is able to avoid (temporarily) the embarrassing problem of picking a gauge in general relativity. As a corollary, one obtains a deep appreciation for the *Nature of Cosmic Time* [7].

**(2) Solution of Constraint Equations.** One may solve the constraint equations of general relativity in a systematic manner, even in a nonlinear setting [4], [8]. The momentum constraint is easy to solve using HJ theory: one simply constructs $\mathcal{S}$ using integrals of the 3-curvature over the entire 3-geometry. The energy constraint may be solved by expanding in powers of the 3-curvature ('spatial gradient expansion').

**(3) Primitive Quantum Theory of Gravity.** Solutions of the HJ equation may be interpreted as the lowest order contribution to the wavefunctional for an expansion in powers of $\hbar$ (semiclassical approximation). One may describe some quantum processes such as the initial 'ground state' of the Universe [9]-[10] or tunnelling through a potential barrier [11].

If one accepts that fluctuations for galaxy formation as well as microwave background anisotropies were generated during an inflationary epoch, then it is imperative that one quantize the gravitational field [9]- [12]. However, quantization of the full gravitational field has proven to be elusive. Several possible forms for a viable theory have been advanced, although there is no general consensus. String theory is the most popular candidate, and its status has been reviewed by many of the speakers in this conference [13]-[14]. The goal here will be more modest. Beginning with the HJ equation, I will be content to consider the semiclassical theory of Einstein gravity. In this way, one follows in spirit the historical development of the theory of atomic spectra. Before the development of the quantum theory in 1926, the semiclassical theory of Bohr and Sommerfeld provided a useful although imperfect description of various atoms.

HJ theory has proven to be a particularly powerful tool for the cosmologist. In this article, I will review our current understanding of the theory. In addition, I will discuss the observational status of three models of cosmological inflation:

(1) inflation with an exponential potential ('power-law inflation') [15], [16] which arises naturally from Induced Gravity [17], [18] or Extended Inflation [19];

(2) inflation with a cosine potential ('natural inflation') [20] where the inflaton is a pseudo-Goldstone boson;

(3) inflation with two scalar fields ('double inflation') [21], [17] where there are two periods of inflation.

## 2  Hamilton-Jacobi Theory for General Relativity

The Hamilton-Jacobi equation for general relativity is derived using a Hamiltonian formulation of gravity. One first writes the line element using the ADM 3+1 split,

$$ds^2 = \left(-N^2 + \gamma^{ij}N_iN_j\right)dt^2 + 2N_i dt\, dx^i + \gamma_{ij}\ dx^i dx^j \ , \qquad (1)$$

where $N$ and $N_i$ are the lapse and shift functions, respectively, and $\gamma_{ij}$ is the 3-metric. Hilbert's action for gravity interacting with a scalar field becomes

$$\mathcal{I} = \int d^4x \left(\pi^\phi \dot\phi + \pi^{ij}\dot\gamma_{ij} - N\mathcal{H} - N^i\mathcal{H}_i\right). \qquad (2)$$

The lapse and shift functions are Lagrange multipliers that imply the energy constraint $\mathcal{H}(x) = 0$ and the momentum constraint $\mathcal{H}_i(x) = 0$.

The object of chief importance is the generating functional

$$\mathcal{S} \equiv \mathcal{S}[\gamma_{ij}(x), \phi(x)]. \tag{3}$$

For each scalar field configuration $\phi(x)$ on a space-like hypersurface with 3-geometry described by the 3-metric $\gamma_{ij}(x)$, the generating functional associates a complex number. The generating functional is the 'phase' of the wavefunctional in the semiclassical approximation:

$$\Psi \sim e^{i\mathcal{S}}. \tag{4}$$

(The prefactor is neglected here although it has important implications for quantum cosmology [24].) The probability functional, $\mathcal{P} \equiv |\Psi|^2$, is given by the square of the wavefunctional. Replacing the conjugate momenta by functional derivatives of $\mathcal{S}$ with respect to the fields,

$$\pi^{ij}(x) = \frac{\delta \mathcal{S}}{\delta \gamma_{ij}(x)} , \qquad \pi^{\phi}(x) = \frac{\delta \mathcal{S}}{\delta \phi(x)} , \tag{5}$$

and substituting into the energy constraint, one obtains the Hamilton-Jacobi equation [22], [23],

$$\begin{aligned}\mathcal{H}(x) = \quad & \gamma^{-1/2} \frac{\delta \mathcal{S}}{\delta \gamma_{ij}(x)} \frac{\delta \mathcal{S}}{\delta \gamma_{kl}(x)} [2\gamma_{il}(x)\gamma_{jk}(x) - \gamma_{ij}(x)\gamma_{kl}(x)] \\ & + \frac{1}{2}\gamma^{-1/2}\left(\frac{\delta \mathcal{S}}{\delta \phi(x)}\right)^2 + \gamma^{1/2} V(\phi(x)) \\ & - \frac{1}{2}\gamma^{1/2} R + \frac{1}{2}\gamma^{1/2}\gamma^{ij}\phi_{,i}\phi_{,j} = 0 , \end{aligned} \tag{6}$$

which describes how $\mathcal{S}$ evolves in superspace. $R$ is the Ricci scalar associated with the 3-metric, and $V(\phi)$ is the scalar field potential. In addition, one must also satisfy the momentum constraint

$$\mathcal{H}_i(x) = -2\left(\gamma_{ik}\frac{\delta \mathcal{S}}{\delta \gamma_{kj}(x)}\right)_{,j} + \frac{\delta \mathcal{S}}{\delta \gamma_{lk}(x)}\gamma_{lk,i} + \frac{\delta \mathcal{S}}{\delta \phi(x)}\phi_{,i} = 0 , \tag{7}$$

which legislates spatial gauge invariance: $\mathcal{S}$ is invariant under reparametrizations of the spatial coordinates. Since neither the lapse function nor the shift function appears in eq.(6) and eq.(7), the temporal and spatial coordinates are *arbitrary*: HJ theory is *covariant*. (Units are chosen so that $c = 8\pi G = \hbar = 1$. The HJ equation for Brans-Dicke gravity has been studied in ref.[25].)

It is quite important that the 3-metric $\gamma_{ij}(x)$ be positive definite otherwise two fatal disasters would occur: (1) the principle of microscopic causality would be violated; (2) points $x$ and $y$ appearing in the generating functional would no longer be space-like, and hence the fields amplitudes $\phi(x)$ and $\phi(y)$ may not be independent —- this would mess up the computation of functional derivatives appearing in eq.(6) and eq.(7).

## 2.1 Partial Reduction of Einstein's Equations

From a real solution $\mathcal{S}$ of the HJ equation and the momentum constraint, one may construct solutions to Einstein's equations. Provided that the fields evolve according to the definitions of the momenta,

$$\left(\dot{\phi} - N^i \phi_{,i}\right)/N = \gamma^{-1/2}\pi^\phi, \tag{8}$$

$$\left(\dot{\gamma}_{ij} - N_{i|j} - N_{j|i}\right)/N = 2\gamma^{-1/2}\pi^{kl}\left(2\gamma_{jk}\gamma_{il} - \gamma_{ij}\gamma_{kl}\right), \tag{9}$$

one may verify that Einstein's equation are indeed satisfied. (Note that | denotes covariant differentiation with respect to the 3-metric.) As a result, HJ theory allows for a partial reduction of Einstein's equations. The lapse and shift appear only in the definitions of the momenta for gravity and matter, eq.(8) and eq.(9). They are arbitrary and they reflect the observer's freedom in choosing his space-time coordinates. For example, if one chooses time hypersurfaces such that the scalar field is uniform, $\phi = t$, then the lapse is given automatically by eq.(8),

$$N^{-1} = \gamma^{-1/2}\frac{\delta \mathcal{S}}{\delta \phi(x)}. \tag{10}$$

Typically one sets the shift to zero, in which case the spatial coordinates of one time slice are projected orthogonally onto the others. In order to complete the determination of the 4-metric as a function of time $\phi$, one need only integrate eq.(9).

## 2.2 Evolution and Observation of a Gravitational System

The above discussion demonstrates how HJ theory enables one to split the analysis of any gravitational system into two stages:

(1) Gauge-independent evolution of the system which is governed by the HJ equation as well as the momentum constraint equation and

(2) Observation of the system which is gauge-dependent (i.e., it depends on arbitrary choices for $N, N_i$).

If the generating functional $\mathcal{S}$ is complex, then one is describing intrinsically *quantum processes*. The freedom in choosing the lapse and shift then reflect the necessity for gauge-fixing. Otherwise, expectation values for physical observables such as the two-point correlation function [1],

$$< \phi(x)\phi(y) > \equiv \int [d\gamma][d\phi]\ \phi(x)\phi(y)\ |\Psi|^2\,, \qquad (11)$$

would be infinite. In the spirit of Dirac quantization [23], gauge-fixing is performed only after one has determined a solution of the HJ equation (or its quantum analog, the Wheeler-DeWitt equation [26]).

# 3 Prototype Hamilton-Jacobi Solution

Until recently, it had been thought that the Hamilton-Jacobi equation for general relativity was intractable. As a result, since the pioneering work of DeWitt [26] and Misner [27] in the 1960's and 1970's, much effort was spent studying minisuperspace models where one effectively truncates superspace: one examines only a finite number of variables describing typically a homogeneous universe. For example, in the context of quantum cosmology, these models were studied by Hartle, Hawking, Page, Vilenkin and others in the 1980's. One hoped to understand some qualitative features of semiclassical gravity from these 'toy models'. However, it is generally agreed that such models represent only the initial stages of a much larger program. For example, in order to appreciate more fully the nature of time, one must incorporate the role of inhomogeneities. After all, a time hypersurface represents the arbitrary manner in which one slices a 4-geometry [2]. Many properties of *full superspace* can be understood by employing a series solution method. The prototype solution, the spatial gradient expansion, was suggested by John Stewart and myself [5]. Effectively, one is able to decompose superspace into an infinite but discrete number of minisuperspaces which are tractable.

## 3.1 Spatial Gradient Expansion

As a first step in solving eq.(6) and eq.(7), one expands the generating functional

$$\mathcal{S} = \mathcal{S}^{(0)} + \mathcal{S}^{(2)} + \mathcal{S}^{(4)} + \ldots , \qquad (12)$$

in a series of terms according to the number of spatial gradients that they contain. The invariance of the generating functional under spatial coordinate transformations suggests a solution of the form,

$$\mathcal{S}^{(0)}[\gamma_{ij}(x), \phi(x)] = -2 \int d^3x \, \gamma^{1/2} H\left[\phi(x)\right] , \qquad (13)$$

for the zeroth order term $\mathcal{S}^{(0)}$. The function $H \equiv H(\phi)$ satisfies the separated HJ equation (SHJE) of order zero [6],

$$H^2 = \frac{2}{3}\left(\frac{\partial H}{\partial \phi}\right)^2 + \frac{1}{3}V(\phi) , \qquad (14)$$

which is an ordinary differential equation. (The term 'separated' describes the fact that the metric variables are absent in eq.(14) —- they have been separated out.) Note that $\mathcal{S}^{(0)}$ contains no spatial gradients. The zeroth order term is an excellent approximation for universes where the wavelength of any inhomogeneities is much larger than the Hubble radius $H^{-1}$. In fact, the notion of *long-wavelength* field is an essential ingredient of any model of cosmic structure formation.

## 3.2 Higher Order Terms

An important simplification occurs for higher order terms, $\mathcal{S}^{(2n)}$, for $n \geq 1$: they are governed by linear partial differential equations of the inhomogeneous type:

$$-2\frac{\partial H}{\partial \phi}\frac{\delta \mathcal{S}^{(2n)}}{\delta \phi} + 2H\gamma_{ij}\frac{\delta \mathcal{S}^{(2n)}}{\delta \gamma_{ij}} + \mathcal{R}^{(2n)} = 0 \ . \tag{15}$$

The remainder term $\mathcal{R}^{(2n)}$ depends on some quadratic combination of the previous order terms [3]. For example, for $n = 1$, it is

$$\mathcal{R}^{(2)} = \frac{1}{2}\gamma^{1/2}\gamma^{ij}\phi_{,i}\phi_{,j} - \frac{1}{2}\gamma^{1/2}R \ , \tag{16}$$

whereas for $n \geq 2$, the remainder, $\mathcal{R}^{(2n)}(x)$, is given by

$$\begin{aligned}\mathcal{R}^{(2n)}(x) = & \ \gamma^{-1/2}\sum_{p=1}^{n-1}\frac{\delta \mathcal{S}^{(2p)}}{\delta \gamma_{ij}(x)}\frac{\delta \mathcal{S}^{(2n-2p)}}{\delta \gamma_{kl}(x)}\left(2\gamma_{jk}\gamma_{li} - \gamma_{ij}\gamma_{kl}\right) \\ + & \ \gamma^{-1/2}\sum_{p=1}^{n-1}\frac{1}{2}\frac{\delta \mathcal{S}^{(2p)}}{\delta \phi(x)}\frac{\delta \mathcal{S}^{(2n-2p)}}{\delta \phi(x)} \ . \end{aligned} \tag{17}$$

In order to compute the higher order terms from eq.(15), one introduces a change of variables, $(\gamma_{ij}, \phi) \to (f_{ij}, u)$:

$$u = \int \frac{d\phi}{-2\frac{\partial H}{\partial \phi}} \ , \quad f_{ij} = \Omega^{-2}(u)\,\gamma_{ij} \ , \tag{18}$$

where the conformal factor $\Omega \equiv \Omega(u)$ is defined through

$$\frac{d\ln\Omega}{du} \equiv -2\frac{\partial H}{\partial \phi}\frac{\partial \ln\Omega}{\partial \phi} = H \ . \tag{19}$$

The equation for $\mathcal{S}^{(2n)}$ simplifies considerably:

$$\left.\frac{\delta \mathcal{S}^{(2n)}}{\delta u(x)}\right|_{f_{ij}} + \mathcal{R}^{(2n)}[f_{ij}(x), u(x)] = 0 \ . \tag{20}$$

Eq.(20) has the form of an infinite dimensional gradient. Before integrating it, I will review some elementary results from potential theory.

## 3.3 Potential Theory

The fundamental problem in potential theory is: given a force field $g^i(u_k)$ which is a function of $n$ variables $u_k$, what is the potential $\Phi \equiv \Phi(u_k)$ (if it exists) whose gradient returns the force field,

$$\frac{\partial \Phi}{\partial u_i} = g^i(u_k) \quad ? \tag{21}$$

Not all force fields are derivable from a potential. Provided that the force field satisfies the integrability relation,

$$0 = \frac{\partial g^i}{\partial u_j} - \frac{\partial g^j}{\partial u_i} = \left[\frac{\partial}{\partial u_j}, \frac{\partial}{\partial u_i}\right] \Phi, \tag{22}$$

(i.e., it is curl-free), one may find a solution which is conveniently expressed using a line-integral

$$\Phi(u_k) = \int_C \sum_j dv_j \; g^j(v_l) \;. \tag{23}$$

If the two endpoints are fixed, all contours return the same answer. In practice, one employs the simplest contour that one can imagine: a line connecting the origin to the observation point $u_k$. Using $s$, $0 \leq s \leq 1$, to parameterize the contour, the line-integral may be rewritten as

$$\Phi(u_k) = \sum_{j=1}^{n} \int_0^1 ds \; u_j \; g^j(su_k) \;. \tag{24}$$

## 3.4 The Nature of Cosmic Time

In solving eq.(20) for the generating functional $\mathcal{S}^{(2n)}$ of order $2n$, one utilizes a line-integral in *superspace*:

$$\mathcal{S}^{(2n)} = -\int d^3x \int_0^1 ds \; u(x) \; \mathcal{R}^{(2n)}[f_{ij}(x), su(x)] \;. \tag{25}$$

For simplicity, the contour of integration was chosen to be a straight line in superspace. As long as the end points are fixed, the line integral is independent of the contour choice which corresponds to a specific time foliation. This property goes a long way in illuminating the nature of time for semiclassical relativity. (Many questions remain concerning the role of time in a quantum setting [28]; because the above arguments have been quite general, I conjecture that line-integrals in superspace will useful for a full quantum description.)

One may verify the integrability of eq.(20) by explicitly computing the commutator [2],

$$\left[\frac{\delta}{\delta u(x)}, \frac{\delta}{\delta u(y)}\right] \mathcal{S}^{(2n)} \equiv \frac{\delta}{\delta u(y)} \mathcal{R}^{(2n)}(x) - \frac{\delta}{\delta u(x)} \mathcal{R}^{(2n)}(y)$$
$$= [\gamma^{ij}(x) \mathcal{H}_j^{(2n-2)}(x) + \gamma^{ij}(y) \mathcal{H}_j^{(2n-2)}(y)] \frac{\partial}{\partial x^i}\delta^3(x-y), \qquad (26)$$

which assumes by induction that $\mathcal{S}^{(2)}, \mathcal{S}^{(4)}, \ldots, \mathcal{S}^{(2n-2)}$ satisfy eq.(20). The 'integrability condition' of potential theory, eq.(22), demands that the commutator (26) vanish. In the above expression, $\mathcal{H}_j^{(2n-2)}$ is the momentum constraint evaluated using the generating functional of order $(2n-2)$:

$$\mathcal{H}_j^{(2n-2)}(x) \equiv -2\left(\gamma_{jk}\frac{\delta \mathcal{S}^{(2n-2)}}{\delta \gamma_{kl}(x)}\right)_{,l} + \frac{\delta \mathcal{S}^{(2n-2)}}{\delta \gamma_{kl}(x)}\gamma_{kl,j} + \frac{\delta \mathcal{S}^{(2n-2)}}{\delta \phi(x)}\phi_{,i}. \qquad (27)$$

We conclude that $\mathcal{S}^{(2n)}$ is indeed integrable provided the term of previous order, $\mathcal{S}^{(2n-2)}$, is invariant under reparametrizations of the spatial coordinates: $\mathcal{H}_j^{(2n-2)} = 0$. In general, the integrability condition for the Hamilton-Jacobi equation follows from the Poisson brackets [29] between the energy densities evaluated at the two spatial points $x$ and $y$:

$$\{\mathcal{H}(x), \mathcal{H}(y)\} = [\gamma^{ij}(x)\mathcal{H}_j(x) + \gamma^{ij}(y)\mathcal{H}_j(y)] \frac{\partial}{\partial x^i}\delta^3(x-y). \qquad (28)$$

Typically, $\mathcal{S}^{(2n)}$ is an integral of terms which contain the Ricci tensor and spatial derivatives of the scalar field [3]. For $n = 1$, one determines that

$$\mathcal{S}^{(2)}[f_{ij}(x), u(x)] = \int d^3x f^{1/2} \left[j(u)\widetilde{R} + k(u)f^{ij}u_{,i}u_{,j}\right]. \qquad (29)$$

$\widetilde{R}$ is the Ricci scalar of the conformal 3-metric $f_{ij}$ appearing in eq.(18). The $u$-dependent coefficients $j$ and $k$ are,

$$j(u) = \int_0^u \frac{\Omega(u')}{2} du' + F, \qquad k(u) = H(u)\,\Omega(u)\,, \qquad (30)$$

where $F$ is an arbitrary constant.

## 3.5 Characteristics of Cosmic Time

The generalization of the spatial gradient expansion to multiple scalar fields is nontrivial [2]. In this case, one employs the method of characteristics for solving the linear partial differential equation that appears. For a single scalar field $\phi$ in a HJ description, it is more or less obvious to use some function of $\phi$ as the integration parameter. In order to facilitate the integration of $\mathcal{S}^{(2)}$ for multiple fields, I recommend using the scale factor, $\Omega \equiv \Omega(\phi_a)$, which is a specific function of the scalar fields. A brief summary follows.

One first solves the SHJE of order zero, eq.(14), describing two scalar fields to find a Hubble function
$$H \equiv H(\phi_1, \phi_2; \widetilde{\phi}_1, \widetilde{\phi}_2) \qquad (31)$$
which depends on two homogeneous parameters $\widetilde{\phi}_1$ and $\widetilde{\phi}_2$. One then makes a change of variables
$$[\phi_1(x), \phi_2(x), \gamma_{ij}(x)] \rightarrow [\Omega(x), e(x), f_{ij}(x)] \qquad (32)$$
where the new variables are found by computing partial derivatives of $H$ with respect to the parameters:

$$\Omega(\phi_a) = \left(\frac{\partial H}{\partial \widetilde{\phi}_1}\right)^{-1/3}, \qquad (33)$$

$$e(\phi_a) = \frac{\partial H}{\partial \widetilde{\phi}_1} \bigg/ \frac{\partial H}{\partial \widetilde{\phi}_2}, \qquad (34)$$

$$f_{ij} = \Omega^{-2}(\phi_a)\,\gamma_{ij}\,. \qquad (35)$$

In terms of the new fields, the generating functional of order two can be computed explicitly:

$$\mathcal{S}^{(2)} = \int d^3x \, f^{1/2} \left[ j(\Omega, e)\widetilde{R} + k_{11}(\Omega, e)\Omega_{;i}\Omega^{;i} + 2k_{12}(\Omega, e)\Omega_{;i}e^{;i} + k_{22}(\Omega, e)e_{;i}e^{;i} \right] . \tag{36}$$

$j$, $k_{11}$, $k_{12}$ and $k_{22}$, are functions of $(\Omega, e)$ which are found to be:

$$j(\Omega, e) = \int_0^\Omega d\Omega' \, \frac{1}{2H(\Omega', e)} + j_0(e) , \tag{37}$$

$$k_{11}(\Omega, e) = \frac{1}{H\Omega} , \tag{38}$$

$$k_{22}(\Omega, e) = \int_0^\Omega d\Omega' \, n(\Omega', e) + k_0(e) , \tag{39}$$

$$k_{12}(\Omega, e) = 0 ; \tag{40}$$

$j_0 \equiv j_0(e)$ and $k_0 \equiv k_0(e)$ are arbitrary functions of $e$; $n(\Omega, e)$ is defined according to

$$n(\Omega, e) = -\frac{1}{2H} \left[ \left(\frac{\partial \phi_1}{\partial e}\right)_\Omega^2 + \left(\frac{\partial \phi_2}{\partial e}\right)_\Omega^2 \right] . \tag{41}$$

Details are described in ref.[2].

# 4 Quadratic Curvature Approximation

In order to describe the fluctuations arising during the inflationary epoch, it is necessary to sum an infinite subset [1], [2], of the terms $\mathcal{S}^{(2n)}$. In this case, one makes an Ansatz which includes all terms which are quadratic in the Ricci tensor $\tilde{R}_{ij}$ of the conformal 3-metric $f_{ij}(x)$:

$$\mathcal{S} = \mathcal{S}^{(0)} + \mathcal{S}^{(2)} + \mathcal{Q} ; \tag{42}$$

here the quadratic functional $\mathcal{Q}$ is

$$\mathcal{Q} = \int d^3x f^{1/2} \left[ \widetilde{R} \, \widehat{S}(u, \widetilde{D}^2) \, \widetilde{R} + \widetilde{R}^{ij} \, \widehat{T}(u, \widetilde{D}^2) \, \widetilde{R}_{ij} - \frac{3}{8} \widetilde{R} \, \widehat{T}(u, \widetilde{D}^2) \, \widetilde{R} \right] , \tag{43}$$

where $\widehat{S}(u, \widetilde{D}^2)$ and $\widehat{T}(u, \widetilde{D}^2)$ are differential operators which are also functions of $u$. $\widehat{S}$ and $\widehat{T}$ describe scalar and tensor fluctuations, respectively. $\widetilde{D}^2$ is the Laplacian operator with respect to the conformal 3-metric. Terms which are cubic and higher are neglected.

## 4.1 Multiple Fields: Quadratic Constant Approximation

Once again, the case for two fields [2] is more complicated (after which the extension to any additional fields is straightforward). One replaces the scalar operator $\widehat{S}$ by a matrix operator $\widehat{S}_{ab}$, $a, b = 1, 2$, which is a function of $\Omega(x)$ and $e(x)$, eq.(33) and eq.(34). The scalar operator $\widehat{S}_{ab}$ is then sandwiched between the vector $[\widetilde{R}, \widetilde{D}^2 e]$ and its transpose in the generalization of eq.(43) to the *quadratic constant approximation*.

# 5 Comparison with Large-Angle Microwave Background Fluctuations and Galaxy Correlations

Using HJ theory, I will compare the cosmological implications of three inflationary models: Model 1 — power-law inflation; Model 2 — natural inflation, and Model 3 — double inflation. All models will be normalized using large angle microwave background anisotropies determined by COBE [30]: $\sigma_{sky}(10^0) = 30.5 \pm 2.7 \mu K$ (68% confidence level).

It is conventional to parametrize the primordial scalar fluctuations arising from inflation by $\zeta$ which is proportional to the metric perturbation on a comoving time hypersurface [6]. The power spectra for $\zeta$ are shown in Figs.(1a), (2a), (3a), for Models 1, 2, 3, respectively. Both power-law inflation and natural inflation yield power spectra $\mathcal{P}_\zeta(k)$ for $\zeta$ that are pure power-laws:

$$\mathcal{P}_\zeta(k) = \mathcal{P}_\zeta(k_0) \left(\frac{k}{k_0}\right)^{n-1} . \tag{44}$$

The spectral index for scalar perturbations is denoted by $n$, and $n = 1$ describes the flat Zel'dovich spectrum. The simplest models arising from inflation are characterized by $n < 1$. The normalization of the spectra differs for the first two models since gravitational waves may contribute significantly to COBE's signal for the power-law inflation model [16]. The primordial power spectrum for double inflation is not a pure power-law. (In Fig.(3a), the primordial fluctuations for inflation with a single field having a quadratic potential is also shown.)

The power spectra $\mathcal{P}_\delta(k)$ for the linear density perturbation $\delta = \delta\rho/\rho$ at the present epoch are shown in Figs.(1b), (2b), (3b). The data points with error bars are determined from the clustering of galaxies [31]. I have assumed that the evolution of the fluctuations is described by the cold-dark-matter transfer function [32] where the present Hubble parameter is taken to be $H_0 = 50$ km s$^{-1}$Mpc$^{-1}$.

For power-law inflation, the best fit is given by $n = 0.9$ (bold curve in Fig.(1b)). However, $n = 0.8$ gives the best fit for natural inflation. From the galaxy data, there is not much difference between the best fits of these two models. One hopes to discriminate these models further when intermediate angle microwave background fluctuations [33] become more precise. Model 3, double inflation model is not particularly attractive since it requires three parameters whereas the previous two models each required one less. For the choice of double inflation parameters advocated by Peter *et al* [34], there is not much advantage over the simpler models of power-law inflation and double inflation [2] (see Fig.(3b)).

## 5.1 Disagreement with Grishchuk

In a previous Erice meeting held in the fall of 1994, Grishchuk [35] stated that the tensor fluctuations arising from virtually all inflation models dominate the contribution to the microwave background anisotropy observed by COBE [30]. I disagreed with this claim [36]. More recently, Deruelle and Mukhanov [37] have demonstrated in careful detail that Grishchuk was in error. Here I wish to clarify the issue further.

The main point of disagreement arises during the computation of scalar pertur-bations and not the tensor perturbations. Once again, the scalar perturbations are parameterized by the variable $\zeta(x)$ which is independent of time when the wavelength of the fluctuation exceeds the Hubble radius. Grishchuk's equation

$$\zeta_{MFB} = \frac{X}{2n^2} \qquad (45)$$

on p.225 of ref.[35] is incorrect. It should read

$$\zeta_{MFB} = \left( X - n^2 \frac{\mu}{a\sqrt{\gamma}} \right) \Big/ (2n^2) \qquad (46)$$

Setting $X = 0$, and then taking $n \to 0$, one finds the standard result that

$$\zeta_{MFB} = -\frac{1}{2} \frac{\mu}{a\sqrt{\gamma}} \qquad (47)$$

approaches a constant which is not equal to zero as $n \to 0$. (An exact solution of the perturbation equation demonstrates this point quite clearly; see, e.g., ref. [1].) Grishchuk erroneously concluded that $\zeta_{MFB} = 0$ because his method of computation was too crude and he neglected the term described above. The scalar contribution to COBE's signal is indeed proportional to $\zeta_{MFB}$, and for many inflation models, it dominates over the tensor component.

# 6 Summary

Hamilton-Jacobi theory for general relativity provides some deep insights into the structure of semiclassical superspace which now far exceeds investigations in homogeneous models. Superspace describes an ensemble of evolving universes, and its complexity strains the imagination. However, the gradient expansion allows one to separate superspace into an infinite but discrete number of manageable pieces which are relatively easy to understand.

The question of time choice in general relativity is a difficult one, particularly for the quantum theory [28]. For semiclassical problems of interest to observational cosmology, one may construct a straightforward covariant formalism which treats all time choices on an equal footing. Different contours of integration in superspace correspond to different time foliations and they all yield the same answer for the generating functional provided that spatial gauge invariance is maintained.

Although not quite perfect, reasonable fits of microwave background anisotropies and galaxy clustering may be obtained by power-law inflation with a spectral index of $n = 0.9$, or by natural inflation with a spectral index $n = 0.8$. If one wishes to pay the price for an additional parameter, double inflation is also adequate.

# 7 Acknowledgments

I thank Prof. Norma Sanchez for organizing yet another beautiful conference in Erice. Many of the Hamilton-Jacobi topics described in this proceedings were developed in collaboration with John Stewart. I thank Alex Vilenkin, Don Page, Werner Israel and Hector DeVega for useful discussions.

This work was supported by NSERC of Canada.

# 8 Figure Captions

**Figs. (1a), (2a), (3a)**: Shown are the power spectra $\mathcal{P}_\zeta(k)$ for zeta, which describes the primordial scalar perturbations arising from inflation. Three plausible models of inflation are considered: (1) power-law inflation, (2) natural inflation and (3) double inflation. The first two models require two parameters: an arbitrary normalization factor and a spectral index $n$, where $n-1$ is the slope of the spectrum in a log-log plot. Double inflation is a three parameter model. For each model, the normalization is fixed by large angle microwave background anisotropies.

**Figs. (1b), (2b), (3b)**: For the present epoch, the power spectra for the linear density perturbation $\delta\rho/\rho$ are shown for the same models of Fig. (a). The data points with error bars are the observed power spectrum derived from galaxy surveys. For power-law inflation, the best fit (bold curve) is obtained with a spectral index of $n = 0.9$, whereas $n = 0.8$ yields the best fit for natural inflation. (Gravitational waves are important for power-law inflation but not for natural inflation.) Using an additional parameter, double inflation also gives a reasonable fit.

**Fig.(1a)**

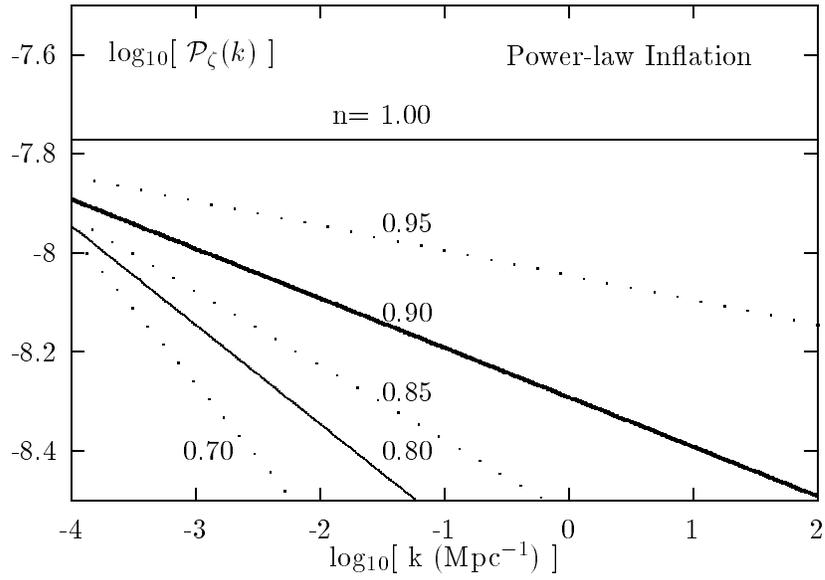

**Fig.(1b)**

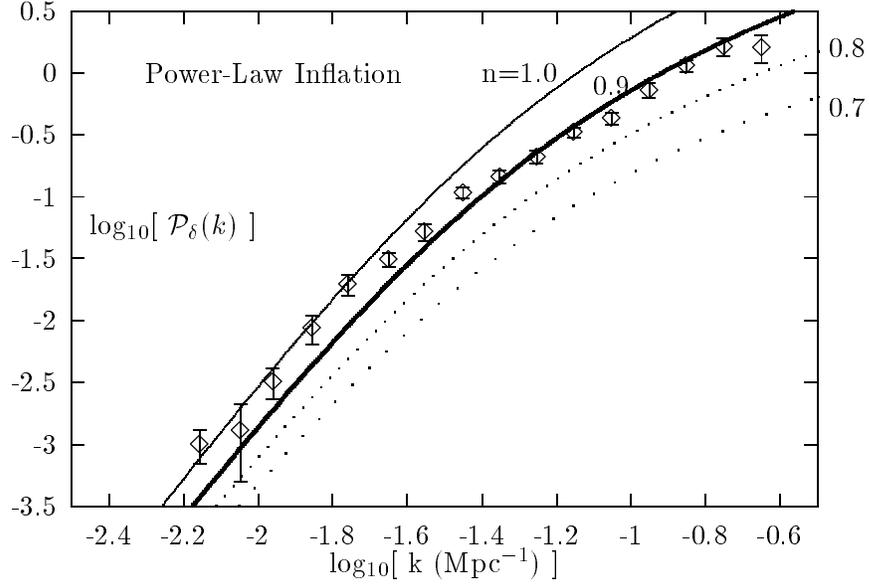

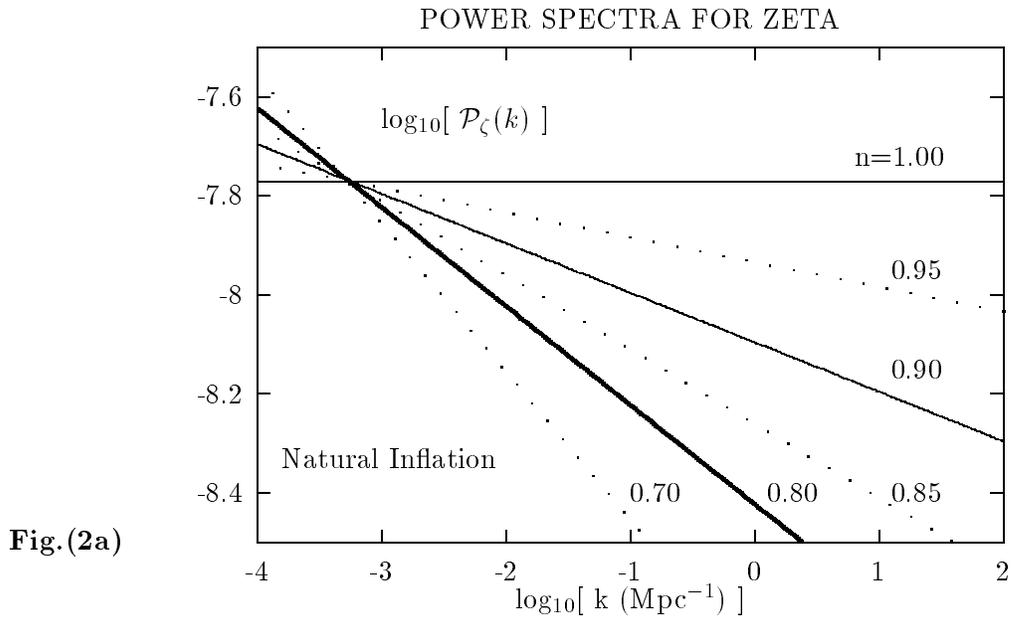

**Fig.(2a)**

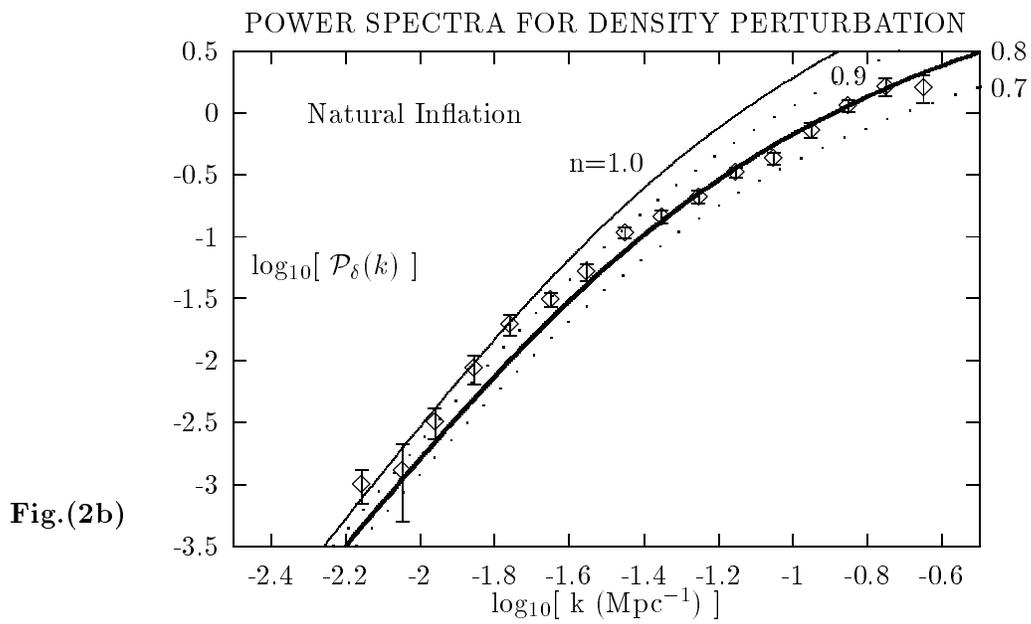

**Fig.(2b)**

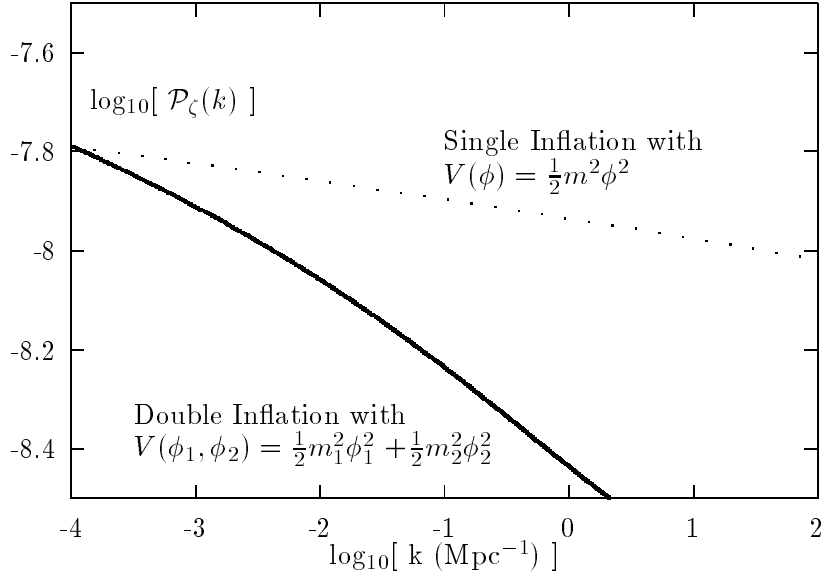

**Fig.(3a)**

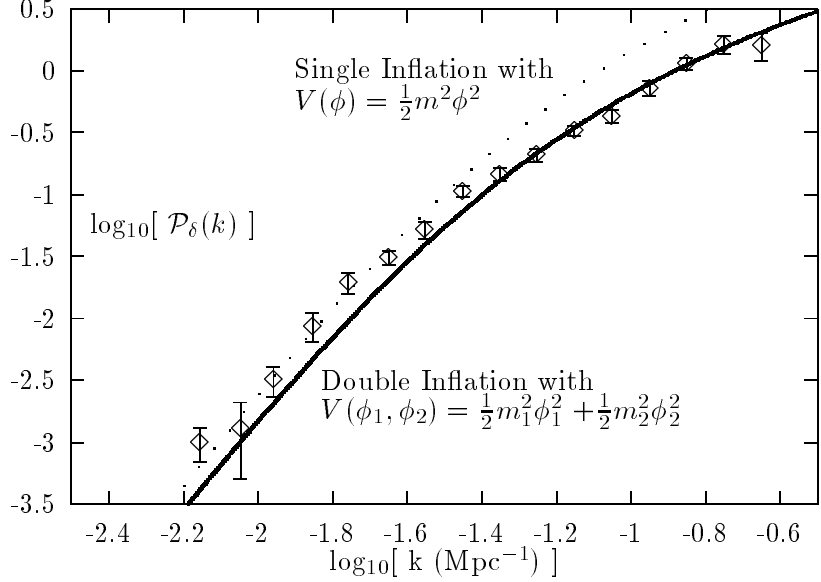

**Fig.(3b)**